# Open Sesame
## The Password Hashing Competition and Argon2

Jos Wetzels

`a.l.g.m.wetzels@student.utwente.nl`

**Abstract.** In this document we present an overview of the background to and goals of the Password Hashing Competition (PHC) as well as the design of its winner, *Argon2*, and its security requirements and properties.

## 1    Background of the Password Hashing Competition (PHC)

The *Password Hashing Competition (PHC)* [1] was an open competition run between 2013 and 2015 by a group of cryptographers that aimed to select one or more password hashing schemes, meeting modern requirements, for recommendation as a (de-facto) standard and ended up selecting the *Argon2* [2] scheme as its winner. Given the ubiquity of passwords as a (primary) means of authentication to a range of services (from web services to mobile, desktop or server systems) or data protection systems (eg. disk encryption solutions), the persistence of password re-use [3] and the (increasing) trend of credential-holding database breaches [4,5,6] password protection is a topic of continuing importance. The initial measure aimed at limiting the impact of database breaches was the usage of cryptographic hash functions for password storage [7] where passwords aren't stored as plaintext but as a hash digest. When users authenticate their password is hashed and compared to the stored digest. The security properties of (cryptographically secure) hash functions [44] aim to make them practically impossible to invert. In this manner a database compromise doesn't result in the immediate disclosure of user credentials but aims to force an attacker to mount a brute-force [9] or dictionary attack [10] on the hashes. Unfortunately such a scheme is vulnerable to space/time trade-off attacks [11] (such as the infamous *rainbow tables* [12]) where an attacker precomputes cryptographic hash functions and stores the result in a lookup table. In this fashion the bulk of the work is performed once, during table generation. Every subsequently attacked hash would simply involve a lookup operation rather than iterative hash computation. The use of *salting* [13] adds a degree of uniqueness to each individual password hash that prevents identical passwords from hashing to identical hashes, slows down brute-force and dictionary attacks (since we cannot compare a candidate against an entire database at once) and mitigates precomputed lookup table attacks. However, hash functions were not designed to be slow (often the contrary), passwords themselves often have very low entropy [14] and technological advances and widespread availability of powerful hardware [15,16] and cracking software leveraging it [17,18] have made fast (bulk-)cracking of password hashes (salted or otherwise) a very feasible affair for many attackers. A solution to this problem roughly comes down to the heuristic of "*slowing down*" calculations to the attacker. A commonly used (but not the only) technique for achieving this is *key stretching* [19] which involves iteratively applying the hash function a tweakable number of times. Using dedicated hardware (eg. GPU, FPGA, ASIC, etc.) [20] implementations, however, cracking such purely '*CPU-bounded*' hashes might still remain within the realm of the feasible for a well-equipped attacker. *Memory-bounded* hash functions [21] were introduced to address this and any modern password hashing scheme will see a mixture of both bounds.

| KDF | 6 letters | 8 letters | 8 chars | 10 chars | 40-char text | 80-char text |
|---|---|---|---|---|---|---|
| DES CRYPT | < \$1 | < \$1 | < \$1 | < \$1 | < \$1 | < \$1 |
| MD5 | < \$1 | < \$1 | < \$1 | \$1.1k | \$1 | \$1.5T |
| MD5 CRYPT | < \$1 | < \$1 | \$130 | \$1.1M | \$1.4k | $1.5 \times 10^{15}$ |
| PBKDF2 (100 ms) | < \$1 | < \$1 | \$18k | \$160M | \$200k | $2.2 \times 10^{17}$ |
| bcrypt (95 ms) | < \$1 | \$4 | \$130k | \$1.2B | \$1.5M | \$48B |
| scrypt (64 ms) | < \$1 | \$150 | \$4.8M | \$43B | \$52M | $6 \times 10^{19}$ |
| PBKDF2 (5.0 s) | < \$1 | \$29 | \$920k | \$8.3B | \$10M | $11 \times 10^{18}$ |
| bcrypt (3.0 s) | < \$1 | \$130 | \$4.3M | \$39B | \$47M | \$1.5T |
| scrypt (3.8 s) | \$900 | \$610k | \$19B | \$175T | \$210B | $2.3 \times 10^{23}$ |

*Fig. 1*: Estimated hardware cost to crack hashed passwords in 1 year as per 2002 [21]

The fallout associated with credential leakage from database breaches involving inadequate hashing schemes is significant [22,23,24] but up until now there has been a low variety of password hashing schemes seeking to address these issues. The only standardized scheme is *PBKDF2* [25] with *bcrypt* [26] and *scrypt* [21] being the main alternatives. And although improving over the alternative of regular hash functions, as illustrated in figure 1, these schemes suffer from several drawbacks. PBKDF2 makes no attempt to minimize GPU/ASIC advantages [27] since it can be implemented rather efficiently with little RAM, *bcrypt* lacks support for tunable memory requirements [28, 8] as well as easily fitting into FPGA designs and *scrypt* does not allow users to only increase time or memory requirements as well as being suboptimal in its defenses against ASICs [29] and TMTO and side-channel attacks [30]. In addition, recent work [20] on GPU- and FPGA-facilitated cracking of *bcrypt* and *scrypt* hashes has shown *scrypt* can be attacked quite efficiently for smaller parameters using GPUs and *bcrypt* can be attacked rather efficiently using FPGAs, as shown in figure 2.

|  | Cost | | Target (CPU) runtime | | | |
|---|---|---|---|---|---|---|
|  | HW | Energy | 1ms | 10ms | 100ms | 1000ms |
| bcrypt | | | | | | |
| − zedboard | $319 | $7.41 | 28.3 H/$s | 2.81 H/$s | 0.303 H/$s | 0.0304 H/$s |
| − GTX 480 | $517 | $759 | 2.25 H/$s | 0.250 H/$s | 0.0264 H/$s | 0.00212 H/$s |
| scrypt | | | | | | |
| − GTX 480 | $517 | $759 | 33.4 H/$s | 1.83 H/$s | 0.0384 H/$s | 0.000287 H/$s |
|  |  |  | (t=1) | (t=2) | (t=8) | (t=4) |

*Fig. 2*: Hashes per dollar-second taking energy and hardware cost for two years into account as per 2015 [20]

As such the PHC was started in order to develop a password hashing scheme capable of meeting the defensive needs of modern applications faced with attackers with modern capabilities. The main focus was on algorithms relevant to most common password hashing scheme applications such as (web) service authentication, key derivation or user login. Apart from security requirements (as discussed in section 2) candidates were evaluated for simplicity (overall scheme clarity, ease of implementation, limited number of internally used primitives or constructions, etc.) and functionality (cost parameter effectiveness, ability to transform existing hashes to different cost settings without password knowledge or possibly even renewed user login, etc.).

In short the security problem addressed by the PHC and the *Argon2* scheme can be formulated as follows: *The PHC and its winner Argon2 seek to present a password hashing scheme that offers the security properties of a 'regular' cryptographically secure hash function while preventing attacks traditionally associated with them and mitigate the effectiveness of both software-optimized and hardware-optimized crackers (thus aiding in eg. reducing the fallout of credential-holding database compromises.*

## 2   Security Requirements of Password-Hashing Schemes

In general [31] password hash cracking involves evaluating multiple candidates (preferably as simultaneous as possible), as fast as possible in order to test as many as possible per given time unit, through the password hashing scheme in question and testing them against one or more target hashes. As such defenses complicating this (slowing down evaluation, reducing opportunities for parallelism, etc.) lie at the heart of password hashing scheme security properties and the overall scheme security is defined by the evaluation of how well it manages to frustrate password cracking efforts. Below follows a list of the core desirable security properties of password hashing schemes as drawn from the PHC requirements [37] and a range of relevant literature [2, 8, 11, 34, 35, 41] based on the types of attacks a hashing scheme is faced with:

- *Cryptographic Security* [8]: The scheme should be cryptographically secure and as such possess the following properties: 1) *Preimage resistance*, 2) *Second preimage resistance* and 3) *collision resistance*. In addition it should avoid other cryptographic weaknesses such as those present in (some) Merkle-Damgård constructions [32] (eg. length extension attacks, partial message collisions, etc. [33]).

- *Defense against lookup table / TMTO Attacks* [11]: The scheme should aim to make TMTO attacks that allow for precomputed lookup table generation, such as Rainbow Tables, infeasible.

- *Defense against CPU-optimized 'crackers'* [34]: The scheme should be '*CPU-hard*', that is, it should require significant amounts of CPU processing in a manner that cannot be optimized away through either software or hardware. As such, cracking-optimized (multi-core) CPU software implementations (eg. written in assembly, testing multiple input sets in parallel) should offer only minimal speed-up improvements compared to those intended for validation ("*slower for attackers, faster for defenders*" [28]).

- *Defense against hardware-optimized 'crackers'* [35]: The scheme should be '*memory-hard*', that is, it should significant amounts of RAM capacity in a manner that cannot be optimized away through eg. TMTO attacks [36]. As such cracking-optimized ASIC, FPGA and GPU implementations should offer only minimal speed up improvements (eg. in terms of *time-area product*) compared to those intended for validation. As noted by Aumasson [28] one of the main scheme design challenges is ensuring minimized efficiency on GPUs, FPGAs and ASICs (in order to minimize benefits of cracking-optimized implementations) and maximized efficiency on general-purpose CPUs (in order to maintain regular use efficiency).

- *Defense against side-channel attacks*: Depending on the use-case (eg. for key derivation or authentication to a device seeking to protect against modification by the device owner) side-channel attacks might be a relevant avenue of attack. Password hashing schemes should aim to offer side-channel resilience. With regards to password hashing scheme security we will focus on security versus the *cache-timing* [38] type of side-channel attacks given the existence of such attacks against the commonly used *scrypt* scheme [30]. The second category of side-channel attacks we will take into consideration are so-called *Garbage Collector Attacks* (GCAs). GCAs have been discussed in literature [41] as an instance of a 'memory leak' attack relevant to password hashing scheme security. GCAs consist of a scenario where an attacker has access to a target machine's internal memory either after termination of the hashing scheme or at some point where the password itself is still present in memory (the so-called *Weak* GCA variant). Both types of attack use their memory observations to find some (intermediate or remnant) value $y$ derived from password input $p$ using function $F$ where testing password candidates $p'$ using $F$ requires significantly less effort compared to the original hashing scheme thus reducing its overall security.

## 3 The *Argon2* Password-Hashing Scheme

*Argon2* [2] is a family of password hashing schemes that was declared winner of the PHC [1]. This section will briefly sketch the *Argon2* scheme and its two major variants: *Argon2d* and *Argon2i*. For the sake of brevity we will omit discussion of its *Server Relief* (SR) and *Client-Independent Update* (CIU) features and similarly consider the (as-of-yet 'non-official') additional scheme variants *Argon2ds* and *Argon2id* out of scope.

Argon2 is optimized for the x86 architecture, (independently) scalable in both time and memory dimensions, supports (single-instance, inner) parallelism and has two major variants: *Argon2d* and *Argon2i*. *Argon2d* is the faster variant using data-dependent memory access (to thwart tradeoff attacks) making it suitable for applications where side-channel attacks are not a threat such as backend servers or cryptocurrencies. It makes a single pass over the memory. *Argon2i* is slower and uses data-independent memory access making it suitable for password hashing and password-based key derivation uses. It makes 3 passes over the memory. *Argon2* takes as its inputs a message $P$ (eg. password), nonce $S$ (eg. salt), parallelism degree $p$, tag length $\tau$, memory size $m$ (where the number of 1024-byte blocks is $m' = \lfloor \frac{m}{4p} \rfloor$), iteration count $t$ (which allows for running time adjustment independently of memory size), secret value $K$ (which serves as optional secret key), associated data $X$, type value $y$ (0 for Argon2d, 1 for Argon2i) and version number $v$ (always $0x10$) giving API $Argon2(P, S, p, \tau, m, t, K, X, y, v)$. Recommended parameter values can be found in the *Argon2* specification.

### 3.1 Operation

*Argon2* has a mode of operation (as illustrated in figure 3) which can iterate for a variable number of passes. It starts by drawing entropy from supplied message $P$ and nonce $S$ which it hashes (together with other parameters) using hash function $\mathcal{H}$ to 512-bit initial hash value $H_0 = \mathcal{H}(p, \tau, m, t, v, y, P', S', K', X')$ where values $P, S, K, X$ are prepended with their lengths ie.: $P' = (len(P), P)$. Next it starts filling memory with $m'$ 8192-bit blocks. Memory is organized as a $p \times q$ matrix where row/lane count is determined by parallelism degree $p$ and column count $q = \frac{m'}{p}$. Blocks are computed as follows, where $H'(X)$ is a variable-length hash function built upon $\mathcal{H}$, $G(X, Y)$ a compression function and block indexes $i'$ and $j'$ are determined by indexing function $\theta(i, j)$:

$$B[i][0] = H'(H_0||i||0), \qquad B[i][1] = H'(H_0||i||1), \qquad 0 \le i < p$$
$$B[i][j] = G(B[i][j-1], B[i'][j']), \qquad 0 \le i < p, \qquad 2 \le j < q$$

If we have multiple passes over the memory (ie.: $t > 1$) the procedure is repeated except for the first two blocks of a lane:
$$B[i][0] = G(B[i][q-1], B[i'][j']), \qquad B[i][j] = G(B[i][j-1], B[i'][j'])$$

In order to enable block computation parallelism memory is partitioned into $S = 4$ vertical *slices* where slice/lane intersections are *segments* of length $\frac{q}{S}$. Segments of the same slice are computed in parallel. After $t$ iterations the final block $B_m$ is computed as the bitwise XOR of the last column and we obtain the output tag $h = H'(B_m)$.

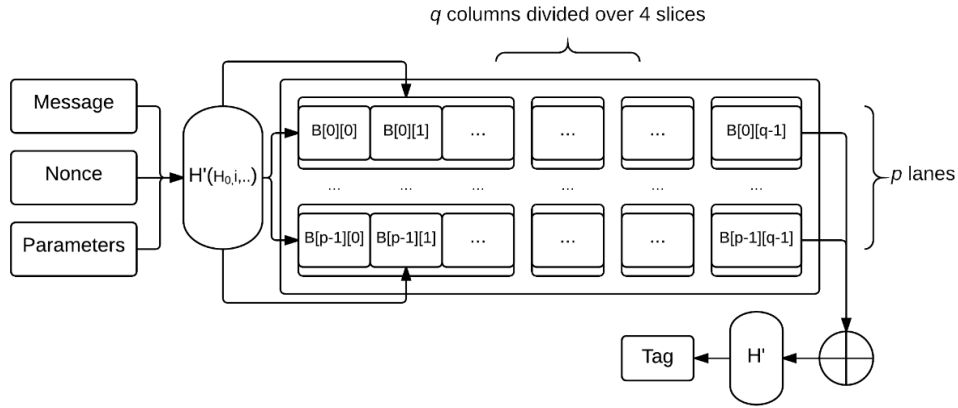

*Fig. 3*: Single-pass *Argon2* with *p* lanes as based upon figure 2 in [2]

### 3.2 Indexing Function $\theta$

During memory filling, the block indexes $i', j'$ are determined by indexing function $\theta$ which differs for *Argon2d* and *Argon2i*. For *Argon2d* we have the data-dependent function $(J_1, J_2) = \theta_D(i,j) = (B[i][j-1]_{0..31}, B[i][j-1]_{32..63})$ where $X_{0..31}$ denotes the first 32 bits of $X$. For *Argon2i* we have the data-independent function $(J_1||J_2) = \theta_I(i,j) = G^2(null_{1024}, \{r||l||s||m'||t||y||i||null_{968}\})$ where $G^2$ is the 2-round version of compression function $G$ (see section 3.4) in counter mode, $null_n$ an *n*-byte all-zero block and the little-endian 8-byte sized pass number $r$, lane number $l$, slice number $s$, total number of memory blocks $m'$, total number of passes $t$, *Argon2* type $y$ and counter $i$ starting from 1 in each segment. The 32-bit values $J_1, J_2$ are then transformed into block indexes $i', j'$ as follows (identical for both *Argon2* versions):

1. Value $l = J_2 \bmod p$ determines lane index from which block will be taken. If $r = s = 0$ (first slice and first pass) then $l$ is set to current lane index.
2. We determine set of candidate indices $\mathcal{R}$ as follows:
    a. If $l$ is current lane then $\mathcal{R}$ includes all not-yet-overwritten blocks computed in this lane excluding $B[i][j-1]$.

b. If $l$ is *not* current lane then $\mathcal{R}$ includes all blocks in last $S - 1 = 3$ segments computed and finished in lane $l$. If $B[i][j]$ is first block of a segment then last block of $\mathcal{R}$ is excluded.
3. We enumerate blocks in $\mathcal{R}$ in order of construction and select the $z^{th}$ block from it where $z = |R| - 1 - (\frac{|R|*((J_1)^2 \,/\, 2^{32})}{2^{32}})$ as our block with index $i', j'$.

### 3.3 Hash Functions $\mathcal{H}$ and $H'$

The hash function $\mathcal{H}$ is defined as $H_{64}$ where $H_x$ is defined as the *BLAKE2b* [42] hash function with $x$-byte output (where $1 \leq x \leq 64$). The derivative variable-length hash function $H'$ is illustrated by figure 4 where $X$ is its input, $V_i$ is a 64-byte block, $A_i$ the corresponding first 32 bytes of that block and $\tau$ the tag length in bytes. Note that the block $V_{r+1}$ is absent if $\tau$ is a multiple of 64.

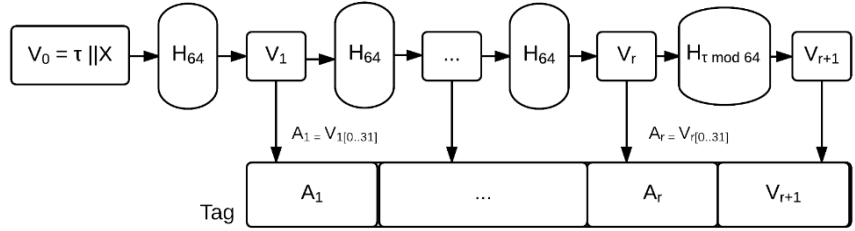

*Fig. 4*: *Argon2* variable-length hash function $H'$

### 3.4 Compression Function G

The compression function $G(X, Y)$ is permutation-based (chosen to avoid the problems associated with iterative compression functions [43]), operates on two 8192-bit input blocks, produces a 1024-bit output block and is built upon the *BLAKE2b* round function $\mathcal{P}$ [42] which in turn operates on (and produces) eight 128-bit blocks and is applied first row-wise and then column-wise as illustrated in figure 5.

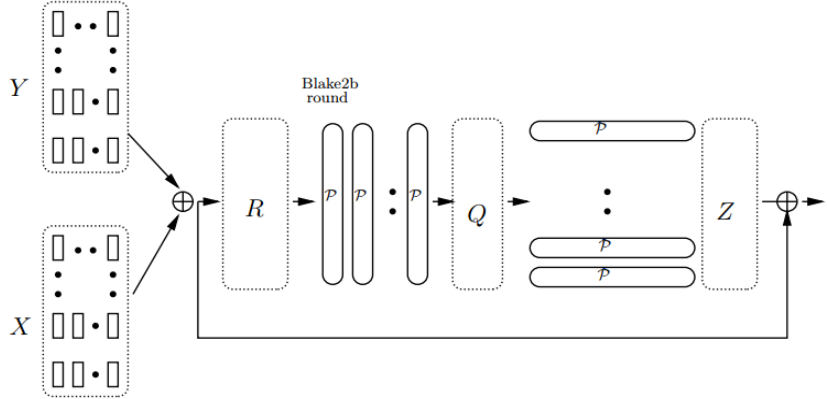

*Fig. 5*: *Argon2* compression function $G$ [2]

## 4  Security Analysis of Argon2

This section discusses how and why the *Argon2* scheme matches the security requirements outlined in section 2. The overall scheme security is determined by its meeting of these individual requirements.

### 4.1  Cryptographic Security

*Collision resistance*: In the initial stage inputs are processed by the hash function $\mathcal{H}$ defined as the *BLAKE2b* function which is considered cryptographically secure. The *Argon2* scheme is claimed [2] (by reductionist proof) to be internally collision resistant even though compression function $G$ is neither claimed to be collision nor preimage resistant. Since the attacker has no control over the inputs of $G$, however, collisions are both very unlikely and uncontrollable. Consider, for example, that an attacker manages to find a collision in $G$ for two arbitrary $m_1, m_2$ such that $G(m_1) = G(m_2)$ then, in order to extend this to a collision for the overall scheme, the attacker would have to find a corresponding preimage for $\mathcal{H}(x)$ (where $x$ is some value determined by 'backtracking' $m_1$ or $m_2$ through the algorithm) which is infeasible considering $\mathcal{H}$ is a cryptographically secure hash function [42]. Furthermore, care is taken to avoid block collisions by initializing starting memory blocks with a counter and making sure the first block in any lane cannot refer (via indexing function $\theta$) to the last block of the previous lane.

*Preimage resistance*: While, as noted above, the internal compression function $G$ is not claimed to be preimage resistant, *Argon2*'s pre-processing of user input with a cryptographically secure hash function $\mathcal{H}$ makes the overall scheme preimage resistant. Consider, for example, an attacker is able to find an $m$ such that $G(m) = h$ for any arbitrary $h$. In order to extend this preimage attack to the overall scheme the attacker would eventually have to find a preimage $n$ such that $\mathcal{H}(n) = m'$ which is infeasible given $\mathcal{H}$ is cryptographically secure [42].

*Second Preimage resistance*: As with the above properties, consider an attacker is able to mount a second preimage attack on compression function $G$ thus being able to find some $m_2$ such that $G(m_1) = G(m_2), m_1 \neq m_2$ for any given $m_1$. In order to extend this attack to the overall scheme the attacker would eventually have to find some $n$ such that $\mathcal{H}(n) = m_2'$ (ie. find a preimage for $\mathcal{H}$).

*Other cryptographic concerns*: There are, as of writing, no mentions of practical cryptographic weaknesses in either the *BLAKE2b* function in specific or the *Argon2* construction in general. The fact that *Argon2* does not rely on Merkle-Damgård constructions anywhere in its design also eliminates the concern of attacks (such as eg. length-extension attacks) specific to that construction [33]. Furthermore, the available cryptanalysis related to *BLAKE2* [62, 63, 64, 65] shows that even attacks targeting reduced-round versions are still purely theoretical and in addition NIST's final report on the SHA-3 competition [66] mentions that *BLAKE* (which is effectively identical at its core to *BLAKE2*) offers a "*very large security margin*" which, given the intensive scrutiny associated with the SHA-3 process, bodes well.

It thus seems safe to conclude that in addition to offering the traditional security properties associated with cryptographic hash functions, the current state-of-the-art seems to indicate no practical cryptographic weaknesses in the *Argon2* scheme.

## 4.2 Defense against lookup table / TMTO attacks

When *Argon2*'s nonces are used (properly), identical messages will produce different tags for different nonces TMTO attacks would require precomputing lookup tables for all possible nonce values which, given the recommended nonce length of 16 bytes, would mean precomputing (and storing) $2^{128}$ lookup tables each of which would be (at least) several gigabytes in size. This, of course, assumes proper nonce generation (preferably using a CSPRNG [45]). If nonce generation happens in a manner where a majority of *Argon2* hashes is generated using the same (predictable) nonce then TMTO attacks could become (more) feasible. It should be noted, however, that *Argon2*'s CPU- and memory-hardness properties (see section 2) allow for setting parameters which would make any precomputation effort in itself (regardless of nonce quality) infeasible to most attackers and as such could mitigate the fallout of such a scenario. Note that this concerns TMTO attacks (such as *Rainbow Tables*) targeting the password hashing scheme in general rather than those which seek to reduce its memory-hardness constraints as mentioned in the section below.

### 4.3 Defense against optimized crackers: CPU-Hardness and Memory-hardness

When defending against optimized crackers it should be noted that though CPU- and memory-hardness might hold for a scheme the essential security offered by it is still a function of its time and memory parameters. As such it remains up to the user to determine their use-case (and corresponding threat model) and adjust parameters accordingly. CPU- and memory-hardness merely aim to guarantee that the scheme makes true on the security claims of its parameters (ie. there are no 'short-cuts').

*CPU-Hardness*: The *Argon2* scheme is optimized for the x86 architecture (by relying on cache and memory organization of recent Intel and AMD processors) and specifically designed to be GPU/FPGA/ASIC-unfriendly. As such pure-CPU implementations are meant to be the most efficient platform for running *Argon2*. Password hashing schemes, however, have deal with optimized CPU-specific cracking implementations which might exploit this built-in platform affinity. The efficiency of any such implementation is bounded by both memory constraints (RAM available for the CPU cores to work with) and the number of cores available to the x86 CPU in question provided there is no (sufficiently efficient) TMTO attack. The *Argon2* designers indicate that multi-core CPUs make support for parallelism attractive (in order to allow for increased bandwidth on the intended defender platform) but as described in section 3, *Argon2* supports a parallelism degree $p$ in a way that prohibits effective TMTO attacks by restricting the internal parallelism to segments (lane/slice intersections). Here the slice count $S = 4$ was chosen because it imposed a low synchronization overhead while simultaneously imposing time-area penalties (see below) on an attacker seeking to perform a TMTO attack. Compared to hardware-specific designs, increasing the number of cores on a complex architecture like x86 scales poorly with respect to the corresponding price: it becomes expensive very soon. Considering the prices of, for example, the 60+ core members of the Intel Xeon Phi x86-based SIMD coprocessor family [46] and the additional investment in memory (even when considering the existing TMTO attack as discussed below) that comes on top of that and combined with the fact that *Argon2*'s parallelism support is TMTO-hardened, it seems safe to conclude that the *Argon2* design holds up well when faced with an (optimized) CPU-only attacker.

*Memory-Hardness*: When defending against hardware-facilitated attackers we can roughly divide them into adversaries with a limited budget, who will opt for GPU and FPGA implementations, and those with significant budgets, who will opt for the more efficient (but also more expensive) ASIC implementations. For various reasons (such as the area occupied by memory, high memory latency in GPUs, higher memory manufacturing costs, etc.) properly designed schemes with intensive memory-use bound the efficiency gain on hardware-facilitated implementations in a way that means they will be neither cheaper nor faster. It is noted in [47] that any scheme which uses more than a few hundred MB of RAM is almost certainly inefficient for GPU- or FPGA-implementations. Consider, for example, the dedicated hash-cracking setup used by the Dutch cyber-security firm Fox-IT [16] which has 4x8GB of RAM and eight AMD Radeon R9 290X GPUs [48] as an example of a professional, well-funded GPU attacker. While the *Argon2* designers mention it remains to be seen how the scheme holds up against GPU crackers with low memory requirements, the scheme's support for multiple GBs of memory usage per tested password instance means that the parallelism (and hence cost-effectiveness) allowed for by even such attacker setups is sharply reduced.

In order to more thoroughly estimate attack-cost, in particular for ASIC-equipped adversaries, and gain an insight into the degree of 'memory-hardness' provided by *Argon2* the time-area product [49, 50] metric is used. The time-area product provides a decent metric since it translates, roughly, into running and production cost and so allows defenders to scale their parameters according to their use-case and threat model. When using the time-area product as a metric what we are looking for is the upper limit to the memory reduction factor an attacker can achieve without increasing the time-area product. If that limit is sufficiently low the scheme is sufficiently memory-hard.

Assume we wish to hash a password using our scheme, then the defender allocates a given amount of time $t$ per password and a certain number of CPU cores before he hashes it using a given amount $M$ of memory. Memory size $M$ translates to some ASIC area $A$ with running time $T$ determined by a mix of computational chain length and memory latency. Then time-area product $AT$ is what we seek to maximize seeing as cracking efficiency gain (in terms of cost and speed) decreases corresponding to $AT$ increase. As per Biryukuv et al. [2] a scheme is called *memory-hard* if, given a cracking implementation requiring reduced memory $\alpha M$ (with $\alpha < 1$) and increased running-time cost $D(\alpha)$ and thus having maximum time-area product gain $g_{max} = \max_\alpha \left(\frac{1}{\alpha D(\alpha)}\right)$, it holds that $D(\alpha) > \frac{1}{\alpha}$ as $\alpha \to 0$. In order to evaluate a scheme's memory-hardness in the face of ASIC-equipped attackers one needs to look at the TMTO attacks that exist against the scheme and determine the maximum time-area product gain the attacker can achieve using these attacks. In general [47], TMTO attacks targeting *data-dependent* memory-hard schemes can reduce the time-area product as long as the additional trade-off running-time cost (related to recomputation tree depth $D$) are smaller than the memory reduction factor, ie. $D(\alpha) \leq \frac{1}{\alpha}$. Attacks targeting *data-independent* schemes are generally efficient until the reduced memory area equals that of the area needed for the multiple recomputation cores.

| $\alpha$ | $\frac{1}{2}$ | $\frac{1}{3}$ | $\frac{1}{4}$ | $\frac{1}{5}$ | $\frac{1}{6}$ | $\frac{1}{7}$ |
|---|---|---|---|---|---|---|
| $C(\alpha)$ | 1.5 | 4 | 20.2 | 344 | 4660 | $2^{18}$ |
| $(D(\alpha)$ | 1.5 | 2.8 | 5.5 | 10.3 | 17 | 27 |

*Fig. 6*: Computation $C$ / Time $D$ penalties for the ranking tradeoff attack on *Argon2*'s data-dependent indexing function [2]

The designers of *Argon2* argue [2] that for both scheme variants *Argon2d* and *Argon2i*, with default number of passes, an ASIC-equipped adversary cannot decrease time-area product if the memory is reduced by a factor of 4 or more, with much higher penalties applying if more passes over the memory are made. Specifically, a *ranking tradeoff attack* [43] exists against both the data-dependent and data-independent variants of *Argon2's* indexing function $\theta$ with results against *Argon2d* illustrated in figure 6. The *Argon2i* variant is more vulnerable to tradeoff attacks due to its data-independent indexing function, hence why it uses 3 passes (rather than 1) by default. It was found by the *Argon2* designers [2] that a memory reduction by a factor of 3 would already mean that the area occupied by the *BLAKE2b* recomputation cores would exceed the area requirements of 1GB of RAM thus making it the upper limit to TMTO effectiveness regardless of scheme variant. In addition to the attacks discussed in [2,43], recent work by Corrigan-Gibs et al. [67] claim tradeoff attacks on *Argon2i* which allow for a memory reduction factor of 5 in the case of single-pass *Argon2i* and a reduction factor of ~ 2.72 for the (recommended) multi-pass variant both without any additional time penalty. These attacks rely on the above discussed fact that for *Argon2i* the block addresses are known in advance and thus blocks which are not used at a given moment can be discarded (until they are overwritten at a later stage) allowing for running the algorithm with smaller memory requirements without corresponding increases in time. In a discussion on the PHC mailing list [68] the idea was proposed to XOR over memory blocks (as the *Lyra2* [69] and *Gambit* [70] schemes do) rather than simply overwriting them which prevents blocks from sitting somewhere 'unused' for some time.

Given that both 'official' variants (with default number of passes) of the *Argon2* scheme allow for a maximum TMTO memory reduction factor of 3, that the scheme comes with recommended memory parameter values of 4GB for backend server authentication, 6GB for encryption key derivation and 1GB for frontend server authentication uses and that users can additionally increase time-area product via the memory parameter and number of passes we can conclude that the scheme's memory-hardness requirement holds up.

### 4.4 Defense against Side-Channel Attacks

*Cache-Timing Attack resistance*: As noted by Biryukov et al. [43] hashing schemes that aim for memory-hardness can be divided into data-independent schemes (which access memory blocks according to a predefined pattern independent of data input) and data-dependent schemes (which access memory blocks according to a function operating on data input). The former allows adversaries to precompute blocks in a just-in-time fashion thus allowing for a reduction in required memory if sufficient computing power is available, effectively constituting a time-memory tradeoff attack. The latter approach prohibits such precomputation but, as is widely documented, data-dependent computations or memory access tend to be vulnerable to side-channel attacks such as cache-timing attacks [51, 52] in the case of (secret or sensitive) data-dependent memory access thus potentially revealing information that could allow an attacker to reduce the computational efforts required to test password candidates. Given that the *Argon2d* variant uses data-dependent memory access in its indexing function $\theta$ (see section 3.2) it is less secure against cache-timing attacks (though their practicality versus *Argon2d* remains to be seen, as mentioned during the PHC selection process [53]) whereas the *Argon2i* variant (which uses data-independent memory access) is not. As such the *Argon2* scheme allows users to adapt depending on their particular use-case and the related security requirements, offering the cache-timing secure *Argon2i* variant to those who need it.

*Garbage Collector Attack (GCA) resistance*: As discussed in section 2, GCAs present a possible threat that could allow an attacker to reduce the security offered by the scheme. By default *Argon2d* makes a single pass over the memory and hence does not overwrite it, leaving it vulnerable to GCAs. *Argon2i*, by contrast, makes three passes over the memory and hence overwrites it twice thus thwarting any kind of GCA since even complete access to all post-computation memory by the attacker would require them to make two passes over the memory to test password candidates. The *Argon2* specifications [2] note that "*if side-channel attacks is a viable threat, enable the memory wiping option in the library call*" without further specifying when and how memory is wiped. From the reference implementation source-code [54] and PHC design discussions [55] it becomes clear, however, that (regardless of *Argon2* variant) if the *clear_memory* flag is set then upon finalization of the scheme [56] the entire working memory is wiped in secure fashion (using a wrapper function for several compiler- and platform-specific secure zero-memory calls). There are also the *clear_password* and *clear_secret* flags which allow for secure wiping of the password $P$ and secret $K$ during initialization. In addition, regardless of flags settings, both the initial hash and the xor-result of the final column are always securely wiped. As such while *Argon2d* is theoretically vulnerable to GCAs the scheme offers options to mitigate this in practice.

### 4.5 Additional Commentary

The *Argon2* family of password hashing schemes is very versatile and can function as password hashing scheme or key derivation function under different conditions and on different defender hardware setups. But this versatility also comes with, in our opinion, its downsides. While the specifications list some recommended parameter values (with regards to memory usage, scheme variant and parallelism degree) for several different use-cases these are rather generic and both hardware setup (available defender CPU and RAM) and use-case (and corresponding threat model) might differ. It can thus be tempting for a software developer, unfamiliar with the details of password hashing schemes and eager to optimize performance, to specify parameter values (in terms of memory usage and number of iteration) which offer insufficient protection in the context of their threat model. In addition the optionality of some memory wiping options (and the little mention they receive in the specifications) might obscure their necessity within certain threat models. In short, the versatility offered by the scheme might give some users 'just enough rope to hang themselves with'. Instead of reducing this versatility however (which might negatively impact widespread adoption of the scheme) it would be preferable if a thorough study was undertaken evaluating different use-cases with different threat models and different hardware setups and distilled recommended parameters for all of these cases as a reference for less cryptographically literate developers.

Interestingly the research that lead to the *Argon2* design [2, 43, 47] showed explicitly the juxtaposition that exists between security against TMTO attacks and side-channel attacks in the design of memory-hard functions with *data-dependent* indexing functions protecting more against the former but being more vulnerable to the latter and vice-versa for *data-independent* functions.

Finally, it should be noted that password hashing schemes in general are no complete mitigation of problems intrinsically associated with passwords as an authentication mechanism. The very nature of passwords, "*easy to remember, hard to guess*", tends to result in people choosing passwords which they *think* are hard to guess but are not [71]. Here it is important to understand that passwords are not generated according to criteria commonly imposed on cryptographic keys: as they are chosen by people, their keyspace and entropy are much lower and they often include predictable elements related to the people who pick them (pet names, date of birth, hobbies, etc.) all of which results in relatively small keyspaces for an attacker to work with. Various mitigations for this problem have been proposed, from using passphrases [75] or '*diceware*' passwords [72] (where passwords consist of a concatenation of multiple individual words chosen at random based on a diceroll) and similar schemes [73] to solutions like *Multi-Factor Authentication (MFA)* where multiple tokens (eg. a password combined with a hardware-based token [74] or code sent over an out-of-band channel such as SMS [75]) are required for authentication. A downside of many of these solutions is that they either lay the responsibility for secure password generation with end-users (eg. diceware, passphrases, etc.) or tend to come with additional development complications and infrastructural cost (eg. MFA).

In addition, passwords or phrases even when composed from the concatenation of multiple (individually easy-to-remember) words are insufficient in the face of clever attackers. Such attackers can generate password candidates on the basis of harvested personal information [76] or use smart, adaptive password cracking strategies or techniques (such as those exhibited in the KoreLogic '*Crack Me If You Can*' contest [57], the use of *Ordered Markov Enumerators* [39] or Markov-modelling based dictionary attacks [40]) to optimize the cracking process through an optimized guessing order [39], reduction of the password space [40] or identification of patterns present in the weakest passwords in a database and extrapolate them to other passwords using masks [58, 59] or thematic dictionaries [60]. Furthermore even if people choose a strong password the omnipresent practice of password reuse [3] remains as a threat. Using the same, strong, password in multiple places means that a compromise of credentials to one (low security) service translates directly to compromise of other (pos-sibly high security) services. To make matters even worse even if an authentication service employs a cryptographically secure hashing scheme the attacker can often simply bypass it by backdooring the login functionality instead [77] and logging the plaintext versions of the password before they are passed to the hashing function. In such a scenario neither strong passwords nor secure password hashing schemes offer any protection and a tendency for password reuse will translate to an impact on other services as well.

In conclusion, while password cracking would definitely be far less successful in the presence of a secure password hashing scheme like *Argon2* it should not be considered a complete silver bullet against problems associated with passwords in general (especially in the face of knowledgeable and well-equipped attackers) and wherever possible authentication schemes ought to be augmented to some form of multi-factor authentication to at least complicate an attacker's efforts in achieving unauthorized authentication.